\documentclass[preprintnumbers,amsmath,amssymb]{revtex4}
\usepackage{amsmath}
\usepackage{amssymb}

\begin{document}

\title{Asymptotic effects in jet
production at high energies\footnote{presented by V.T.K. at QUARKS'2008, Zagorsk, Russia, May 25-31, 2008}}

\author{V.B.~Gavrilov$^\&$, V.T.~Kim$^\sharp$, A.A.~Krokhotin$^\&$, G.B.~Pivovarov$^\ddagger$ and G.B.~Safronov$^\&$}
\affiliation{
$\&$ Institute of Theoretical and Experimental Physics,
Moscow, 117218 Russia\\
$\sharp$ St. Petersburg Nuclear Physics Institute,
Gatchina, 188300 Russia\\
$\ddagger$ Institute for Nuclear Research,
Moscow, 117312 Russia}



\begin{abstract}
Monte Carlo event simulation with BFKL evolution is discussed.
We report current status of a Monte Carlo event generator ULYSSES with
BFKL evolution implemented.
The ULYSSES, based on Pythia Monte Carlo generator, would help
to reveal BFKL effects at LHC energies.
In particular, such an observable as dijet K-factor can serve
as a source of BFKL dynamics at the LHC, and it would also help to search for new physics.
\end{abstract}


\maketitle

With advent of the Large Hadron Collider  (LHC) at CERN we have an opportunity to probe
the Standard Model far beyond the explored domains.
QCD is an essential ingredient of the Standard Model,
and it is well tested in  hard processes when transferred momentum
is of the order of the
total collision energy (Bjorken limit: $Q^2 \sim s \rightarrow \infty$).
The cornerstones of perturbative QCD
at this kinematic regime (QCD-improved parton model)are
factorization of inclusive hard processes \cite{Amati}
and Gribov--Lipatov--Altarelli--Parisi--Dokshitzer
(GLAPD) evolution equation \cite{GLAPD}. They provide
a basis for the successful
QCD-improved parton model. The factorization theorem  \cite{Amati}
for inclusive hard  processes
ensures that the inclusive cross section  factorizes
into partonic subprocess(es)
and parton distribution function(s).
The GLAPD-evolution equation governs the $\log Q^2$-dependence
(at $Q^2 \rightarrow \infty$)
of the
inclusive hard process cross-sections at fixed scaling variable $x \sim Q^2/s$.

Another kinematic domain that is very important
at high-energy is given by the
Balitsky--Fadin--Kuraev--Lipatov
(BFKL) limit
\cite{FKL,BL78,Lip97,GLR},
or QCD-Regge limit,
whereby at fixed $Q^2 \gg \Lambda_{QCD}^2$, $s \rightarrow \infty $.
In the BFKL limit, the BFKL evolution in the leading-log approximation (LLA)
governs $\log(1/x)$ evolution (at $x \rightarrow 0$)
of inclusive processes.

One of the key BFKL features  is a relation of the highest eigenvalue, $\omega^{max}$, of the BFKL
equation
\cite{FKL,BL78,Lip97,GLR}
[with = to] the intercept of the Pomeron, which in turn governs
the high-energy asymptotics of the total cross-sections: $\sigma \sim
(s/s_0)^{\alpha_{I \negthinspace P}-1} = (s/s_0)^{\omega^{max}}$, where
the Regge parameter $s_0$ defines the approach to
the asymptotic regime.
The BFKL Pomeron intercept in the LLA turns out to be rather large:
$\alpha_{I \negthinspace P} - 1 =\omega_{LLA}^{max} =
12 \, \log2 \, ( \alpha_S/\pi )  \simeq 0.54 $ for
$\alpha_S=0.2$;
hence, it is very important to take into account
 NLLA corrections \cite{FL,BFKLP,Ciafaloni99,Thorne99,Ball99} to the BFKL.
Note that the BFKL evolution
in the next-to-leading-log approximation (NLLA)
\cite{FL,BFKLP,KLP},
unlike the LLA BFKL
\cite{FKL,BL78,Lip97},
includes GLAPD evolution
with the running coupling constant of the leading-order (LO) GLAPD,
$\alpha_S(Q^2) = 4 \pi / \beta_0
\log(Q^2/\Lambda_{QCD}^2)  $.

One of the striking features of the NLLA BFKL analysis \cite{BFKLP}
is that the NLLA value for
the intercept of the BFKL Pomeron, improved by the BLM procedure \cite{BLM},
has a very weak dependence on the gluon virtuality $Q^2$:
$\alpha_{I \negthinspace P} - 1 =\omega_{NLLA}^{max}   \simeq$ 0.13 -- 0.18
at $Q^2 = 1$ -- 100 GeV$^2$.
This agrees with the conventional Regge theory where
one expects  universal intercept of the Pomeron without any $Q^2$-dependence.
The value of NNLA BFKL Pomeron intercept \cite{BFKLP} becomes compatible with
the available data on hard QCD Pomeron. So, NLLA BFKL  approach \cite{BFKLP,KLP} posses
all basic features of BFKL, but includes also running coupling effects and
realistic value of the hard Pomeron intercept.

Therefore, the BFKL and especially the NLLA BFKL
\cite{FL,BFKLP,KLP}
are anticipated to be
important tools for exploring the high-energy limit of QCD.

It should be stressed that in contrast to the GLAPD, BFKL dynamics involves
parton distributions unintegrated over $k_t$. So, to reveal BFKL effects one needs to deal, e.g., with
parton (jet) production. For jet production in the LLA BFKL within $k_t$-factorization \cite{Catani91}
one can use effective Feynman-like rules for inclusive jet cross sections \cite{KP}.

However, extremely sophisticated design of contemporary high-energy detectors requires
from theory and phenomenology a Monte Carlo event generator implementation.
Widely used Monte Carlo event generator Pythia \cite{Pythia} contains basically
GLAPD-evolution. To incorporate BFKL-evolution Pythia has been modified
in parton shower kernel and parton distribution functions.

For Monte Carlo event simulations of GLAPD evolution it is
convenient to use the Dokshitzer-Diakonov-Troyan (DDT) \cite{DDT} representation:
\begin{equation} \label{eq:DGLAP1}
  \frac{\partial\, a(x,k_t^2)}{\partial \log k_t^2}  =
  \frac{\alpha_S(k_t^2)}{2 \pi}\,\sum_{b}\,\int_x^1\! d{z}\,P_{ab }(z)\,b
  \left (\frac{x}{z}, k_t^2 \right) - \frac{a(x,k_t^2)}{T_a(k_t^2,\mu^2)}\,
  \frac{\partial \,T_a (k_t^2,\mu^2)}{\partial \log k_t^2}, \nonumber
\end{equation}
where $P_{ab }(z)$ are kernels (splitting functions) of GLAPD equation for partons $a$ and $b$,
\begin{equation} \label{eq:norm}
  a(x,\mu^2) = \int^{\mu^2}\!d{k_t^2}\,f_a(x,k_t^2,\mu^2),
  \,\,\, a=g,q,\bar{q} \nonumber
\end{equation}
  is GLAPD parton distribution for parton $a$, $f_a(x,k_t^2,\mu^2)$
is so-called unintegrated parton distribution used in BFKL approach
and
\begin{equation} \label{eq:Sudakov}
  T_a (k_t^2,\mu^2) \equiv \exp \left (-\int_{k_t^2}^{\mu^2}\! d{\kappa_t^2}\,
  \frac{\alpha_S(\kappa_t^2)}{2\pi}\,\sum_{b}\,\int_0^1\! d{\zeta}\;\zeta
  \,P_{b  a}(\zeta ) \right ) \nonumber
\end{equation}
is QCD analog for Sudakov form factor (DDT form factor \cite{DDT}), which defines the probability
not to emit a gluon.

In terms of unintegrated parton distribution the GLAPD equation
can be represented as:
\begin{eqnarray}
\label{eq:UPDF}
 \,\,
   f_a(x,k_t^2,\mu^2) & \equiv & \frac{\partial}{\partial \log k_t^2}\left[\,a(x,k_t^2)
  \,T_a(k_t^2,\mu^2)\,\right]\nonumber \\ \,\,\,\,\,\, &=&
  T_a(k_t^2,\mu^2)\,\frac{\alpha_S (k_t^2)}{2\pi}\,\sum_{b }\,\int_x^1\! dz
  \,P_{ab }(z)\,b \left (\frac{x}{z}, k_t^2 \right) \nonumber
\end{eqnarray}
Then, a unified BFKL-GLAPD equation can be presented in the following way \cite{Ryskin}:
\begin{eqnarray}
\label{eq:a14} && f_g (x, k_t^2, \mu^2)
 =  T_g (k_t, \mu) \: \frac{\alpha_S (k_t^2)}{2 \pi} \left
\{\int_x^{\mu/(\mu + k_t)} \: dz \:  \int^{k_t^2} \:
\frac{dk_t^{\prime 2}}{k_t^{\prime 2}} \: \left [\bar{P}(z) \:  h_g
\left (\frac{x}{z}, k_t^{\prime 2} \right ) \right .  \right .
\nonumber \\ & & \nonumber \\ & & + \; \left . P_{gq}(z) \: \sum h_q
\left (\frac{x}{z}, k_t^{\prime 2} \right ) \right ] \:  + \: 2N_C
\:  \int_x^{\mu/(\mu + k_t)} \:  \frac{dz}{z} \:  \int \:  \frac{d^2
q}{\pi q^2} \nonumber \\ & & \nonumber \\ & & \left . \times \left [
\frac{k_t^2}{k_t^{\prime 2}} \:  h_g \left ( \frac{x}{z},
k_t^{\prime 2} \right ) \:  - \:  \Theta (k_t^2 - q^2) \:  h_g \left
(\frac{x}{z}, k_t^2 \right ) \right ] \right \}, \, \bar{P}(z) \; =
\; P_{gg} (z) \: - \: \frac{2N_C}{z}
\end{eqnarray}
The Eq.~(\ref{eq:a14}) is similar to Ciafaloni-Catani-Fiorani-Marchesini (CCFM) \cite{CCFM} equation, which is a version of BFKL equation with imposed angle ordering.
\begin{center}
\begin{figure}[htb]
\vspace*{8.0 cm}
\begin{center}
\includegraphics{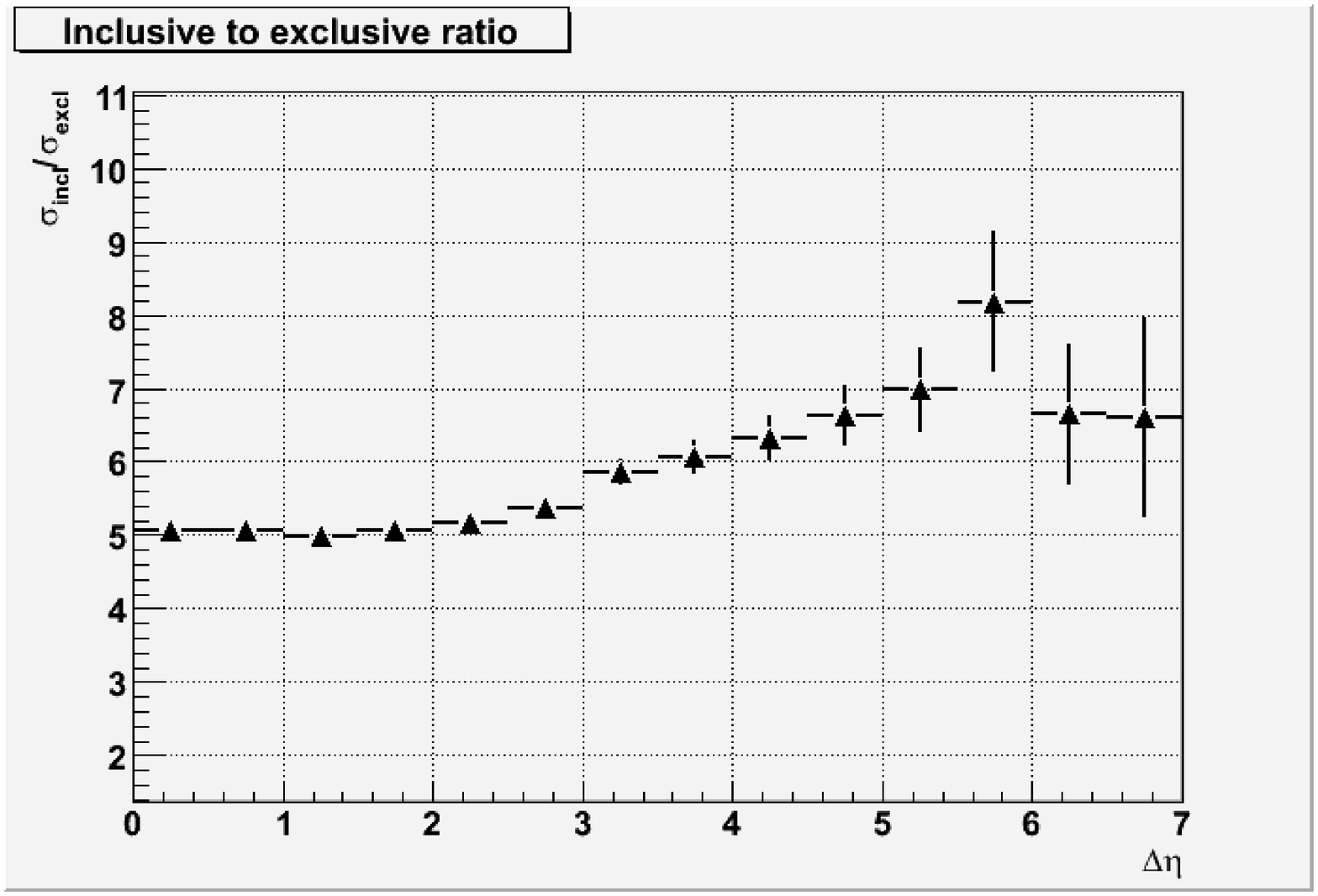}
\caption{Dijet K-factor at LHC: inclusive dijet cross section over only-two-jet cross section ratio.
The results are shown with full CMS detector simulation (arbitrary normalization) for luminosity 20 pb$^{-1}$.}
\end{center}
\label{fig:fig1}
\end{figure}
  \end{center}
Monte Carlo event generator Ulysses \cite{Ulysses} contains an implementation of the Eq.~(\ref{eq:a14}) into Pythia \cite{Pythia} parton shower simulation
and embedded BFKL unintegrated parton distributions. Preliminary results for dijet K-factor at CMS detector of the LHC are shown in Fig.~1, where dijet K-factor for jets with $E_t > E_t^{min}=$ 60 GeV is defined as ratio of inclusive dijet cross section to "exclusive" one. "Exclusive" dijet event means the event when only two jets with $E_t$  above $E_t^{min}$
are available ("Born" cross section).

To summarize, BFKL Monte Carlo should help to look at LHC for BFKL dynamics as a new feature of asymptotic QCD.
It also has a potential for search of new physics, e.g., graviton production in trans-Planckian regime \cite{Guidice}.

V.T.K.  thanks the organizers of the QUARKS'2008 for their warm hospitality.
This work was supported in parts by RFBR grant No. 08-02-01184a, and
RF President grants Nos. 1616.2008.2 and 378.2008.2.



\begin{thebibliography}{11}
\expandafter\ifx\csname natexlab\endcsname\relax\def\natexlab#1{#1}\fi
\expandafter\ifx\csname bibnamefont\endcsname\relax
  \def\bibnamefont#1{#1}\fi
\expandafter\ifx\csname bibfnamefont\endcsname\relax
  \def\bibfnamefont#1{#1}\fi
\expandafter\ifx\csname citenamefont\endcsname\relax
  \def\citenamefont#1{#1}\fi
\expandafter\ifx\csname url\endcsname\relax
  \def\url#1{\texttt{#1}}\fi
\expandafter\ifx\csname urlprefix\endcsname\relax\def\urlprefix{URL }\fi
\providecommand{\bibinfo}[2]{#2}
\providecommand{\eprint}[2][]{\url{#2}}

\bibitem{Amati}
D. Amati, R. Petronzio and G. Veneziano, Nucl.~Phys. {\bf B140},
54 (1978); {\it ibid.} {\bf B146}, 29 (1978); \\
S.~B. Libby and G. Sterman, Phys.~Rev. {\bf D18}, 3252 (1978); \\
A.~H. Mueller, Phys.~Rev. {\bf D18}, 3705 (1978); \\
A.~V. Efremov and A.~V. Radyushkin,
JINR-E2-11725, 11726, 11849, Dubna (1978),
Teor.~Mat.~Fiz. {\bf 44}, 17, 157, 327 (1980)
[J. Theor. Math. Phys. {\bf 44}, 573, 664, 774 (1981)]; \\
R.~K.~Ellis, H.~Georgi, M.~Machachek, H.~D.~Politzer and
G.~G.~Ross, Nucl.~Phys. {\bf B152}, 285 (1979); \\
J.~C. Collins, D.~E. Soper and G.~Sterman,
Nucl.~Phys. {\bf B261}, 104 (1985); \\
G.~T. Bodwin, Phys. Rev.  {\bf D31}, 2616 (1985); {\it ibid.} (E)
{\bf D34}, 3932 (1986); \\
for a review, see J.~C. Collins, D.~E. Soper and G.~Sterman, in:
{\it Perturbative QCD}, ed. A.~H. Mueller (World Scientific, Singapore, 1989)
p. 1.

\bibitem{GLAPD}
V.~N. Gribov, L.~N. Lipatov, Yad.~Fiz. {\bf 15}, 781, 1218 (1972)
[Sov. J. Nucl. Phys. {\bf 15}, 438, 675 (1972)]; \\
L.~N. Lipatov, Yad. Fiz. {\bf 20}, 181 (1974)
[Sov. J. Nucl. Phys. {\bf 20}, 94 (1975)]; \\
G. Altarelli and G. Parisi, Nucl. Phys. {\bf B126}, 298 (1977); \\
Yu.~L. Dokshitzer,  ZhETF {\bf 73}, 1216 (1977)
[Sov. Phys. JETP {\bf 46}, 641 (1977).



\bibitem{FKL}
E.~A.~Kuraev, L.~N.~Lipatov and V.~S.~Fadin,
Phys. Lett. B{\bf 60}, 50 (1975); ZhETF
{\bf 71}, 840 (1976) [Sov. Phys. JETP {\bf 44}, 443 (1976)];
{\it ibid.} {\bf 72}, 377 (1977) [{\bf 45}, 199 (1977)]; \\
L.~N. Lipatov, Yad. Fiz. {\bf 23}, 642 (1976)
[Sov. J. Nucl. Phys. {\bf 23}, 338 (1976)]; ZhETF {\bf 90}, 1536 (1986)
[Sov. Phys. JETP {\bf 63}, 904 (1986)].

\bibitem{BL78}
I.~I.~Balitsky and L.~N.~Lipatov, Yad. Fiz. {\bf 28}, 1597 (1978)
[Sov. J. Nucl. Phys. {\bf 28}, 822 (1978)].

\bibitem{Lip97}  L.~N.~Lipatov, Phys.~Rep. C{\bf 286}, 131 (1997).

\bibitem{GLR}
L.~V. Gribov, E.~M. Levin and M.~G. Ryskin, Phys. Rep. C{\bf 100}, 1 (1983).

\bibitem{FL}
V.~S.~Fadin and L.~N.~Lipatov, Phys. Lett. B{\bf 429}, 127 (1998); \\
 G.~Camici and M.~Ciafaloni, Phys. Lett. B{\bf 430}, 349 (1998);
 and references therein.



\bibitem{BFKLP}  S.~J.~Brodsky, V.~S.~Fadin, V.~T.~Kim, L.~N.~Lipatov
and G.~B.~Pivovarov, Pisma~ZhETF~{\bf 70}, 161 (1999)
[JETP~Lett. {\bf 70}, 155 (1999)], hep-ph/9901229.

\bibitem{Ciafaloni99}
M.~Ciafaloni, D.~Colferai, and G.~P.~Salam,
Phys. Rev. D{\bf60} (1999) 114036;\\
M.~Ciafaloni and D.~Colferai, Phys. Lett. B{\bf452} (1999) 372.

\bibitem{Ball99}
G.~Altarelli, R.~D.~Ball, and S.~Forte,
Nucl. Phys. B{\bf575} (2000) 313.


\bibitem{Thorne99}
R.~S.~Thorne, Phys. Rev. Phys. Rev. D{\bf60}  (1999) 054031;
{\it ibid.} D{\bf64} (2001) 074005.

\bibitem{KLP} V.~T.~Kim, L.~N.~Lipatov and G.~B.~Pivovarov,
IITAP-99-013 Ames (1999), hep-ph/9911228; IITAP-99-014 Ames (1999), hep-ph/9911242.


\bibitem{BLM}  S.~J.~Brodsky, G.~P.~Lepage and P.~B.~Mackenzie,
Phys.~Rev. D{\bf 28}, 228 (1983).

\bibitem{Catani91}
S. Catani, M. Ciafaloni and F. Hautmann,
Nucl. Phys. B{\bf 366}, 135 (1991);
J.~C. Collins and R.~K. Ellis, Nucl. Phys. B{\bf 360}, 3 (1991);
E.~M. Levin, M.~G. Ryskin, Yu.~M. Shabelsky  and A.~G. Shuvaev,
Yad. Fiz. {\bf 54} 1420 (1991) [Sov. J. Nucl. Phys. {\bf 54}, 867 (1991)].

\bibitem{KP}
 V.~T.~Kim, G.~B.~Pivovarov,
Phys. Rev. D{\bf54}, 725 (1996).

\bibitem{Pythia}
T.~Sjostrand, S.~Mrenna, P.~Skands, JHEP {\bf 0605}, 026 (2006).

\bibitem{DDT}
Yu.~L.~Dokshitzer, D.~I.~Diakonov and S.~I.~Troyan,
Phys. Rep. C{\bf 92}, 90 (1980);

\bibitem{Ryskin}
M.~A.~Kimber, J.~Kwiecinski, A.~D.~Martin and A.~M.~Stasto,
Phys.Rev. D{\bf 62}, 4006 (2000); \\
M.~A.~Kimber, A.~D.~Martin and M.~G.~Ryskin,
Phys.Rev. D{\bf 63}, 114027 (2001).

\bibitem{CCFM}
M.Ciafaloni, Nucl. Phys. {\bf B296}, 49 (1988); \\
S.~Catani, F.~Fiorani, G.~Marchesini, Phys. Lett. B{\bf 234}, 339 (1990);
Nucl. Phys. B{\bf 336}, 18 (1990); \\
G.~Marchesini, Nucl. Phys. B{\bf 445}, 49 (1995)

\bibitem{Ulysses}
V.~T.~Kim, A.~A.~Krokhotin and G.~B.~Pivovarov, in prepration.

\bibitem{Guidice}
G. F. Giudice, R. Rattazzi, J. D. Wells,
Nucl.Phys. B{\bf 544}, 3 (1999); Nucl. Phys. B{\bf 630}, 293 (2002).


\end{thebibliography}
\end{document}